\documentclass[aps,prb,twocolumn,superscriptaddress,nofootinbib,longbibliography]{revtex4-1}
\usepackage{amsmath}
\usepackage{amssymb}
\usepackage{graphicx,subfigure}
\usepackage{mathrsfs}
\usepackage{ntheorem}
\usepackage[T1]{fontenc}
\usepackage{bm}
\usepackage{subfigure}
\usepackage[bookmarks=false]{hyperref}
\usepackage{xcolor}
\usepackage{ulem}
\hypersetup{colorlinks=true,citecolor=blue,
linkcolor=blue,urlcolor=blue,pdfstartview=FitH,
bookmarksopen=true}

\newcommand{\ket}[1]{|#1\rangle}

\newcommand{\inp}[2]{\langle #1|#2\rangle}

\newcommand{\ud}{\mathrm{d}}

\newcommand\st{\bgroup\markoverwith
{\textcolor{red}{\rule[0.5ex]{2pt}{0.4pt}}}\ULon}
\newcommand\ul{\bgroup\markoverwith
{\textcolor{blue}{\rule[-0.5ex]{2pt}{0.4pt}}}\ULon}

\begin{document}

\title{Revealing many-body effects on interband coherence through adiabatic charge pumping}

\author{Sen Mu}
\affiliation{Department of Physics, National University of Singapore, Singapore 117542}
\author{Da-Jian Zhang}
\affiliation{Department of Physics, National University of Singapore, Singapore 117542}
\author{Longwen Zhou}
\email{zhoulw13@u.nus.edu}
\affiliation{Department of Physics, College of Information Science and Engineering, Ocean University of China, Qingdao, 266100, China}
\author{Jiangbin Gong}
\email{phygj@nus.edu.sg}
\affiliation{Department of Physics, National University of Singapore, Singapore 117542}

\date{\today}

\begin{abstract}

The adiabatic charge pumping of a non-equilibrium state of spinless fermions in a one-dimensional lattice is investigated, with an emphasis placed on its usefulness in revealing many-body interaction effects on interband coherence.  For a non-interacting system, the pumped charge per adiabatic cycle depends not only on the topology of the occupied bands but also on the interband coherence in the initial state.  This insight leads to an interesting opportunity for quantitatively observing how quantum coherence is affected by many-body interaction that is switched on for a varying duration prior to adiabatic pumping.  In particular, interband coherence effects can be clearly observed by adjusting the switch-on rates with different adiabatic pumping protocols and by scanning the duration of many-body interaction prior to adiabatic pumping. The time dependence of single-particle interband coherence in the presence of many-body interaction can then be examined in detail.  As a side but interesting result, for relatively weak interaction strength, it is found that the difference in the pumped charges between different pumping protocols vanishes if a coherence measure defined from the single-particle density matrix in the sublattice representation reaches its local minima.  Our results hence provide an interesting means to quantitatively probe the dynamics of quantum coherence in the presence of many-body interaction (e.g., in a thermalization process).


\end{abstract}

\maketitle


\section{Introduction}

Recent experiments on ultracold atoms platforms have made it possible to manipulate and control non-equilibrium dynamics of quantum many-body systems\cite{Greiner2002a,Greiner2002,Greiner2005,Hofferberth2007,Zhang2012,Hart2015,Song2018}. A practical way to excite a many-body system from its equilibrium ground state to a non-equilibrium state is through quantum quenches, i.e. a sudden change of parameters in its Hamiltonian\cite{Manmana2007,Kollath2007,Moeckel2008,Eckstein2009}.
The quantum evolution dynamics of the non-equilibrium states can remain coherent for long time due to the almost perfect isolation of the atoms from their environment\cite{Jaksch2005,Bloch2008,Cazalilla2011}. It is expected that a non-equilibrium initial state evolves towards thermal equilibrium for generic isolated quantum systems unless many-body localization emerges \cite{Srednicki1994,Cramer2008,Rigol2008,Barthel2008,Banuls2011,Cassidy2011,Polkovnikov2011,Nandkishore2014,DAlessio2016,Abanin2018}. Yet, theoretical description for several aspects of this process remains challenging, since many well-established theoretical methods for equilibrium systems fail in the non-equilibrium regime\cite{Rigol2009}, e.g. How does the system finally thermalize, or is that possible to extract any memory of the initial state? Explicitly how does the quantum coherence of a non-equilibrium initial state evolve? These are important questions because quantum coherence is a fundamental feature of quantum mechanics and underpins a plethora of fascinating phenomena in various areas of physics\cite{Scully1991,Albrecht1994,Giovannetti2006,Lostaglio2015,Narasimhachar2015} and even biology \cite{Engel2007,Collini2010,Lambert2013}. Therefore, the detailed dynamics of the quantum coherence in thermalization processes is worth investigating.

Thouless pumping demonstrates a profound concept in condensed matter physics, as it establishes a deep connection between the band topology and quantum transport\cite{Thouless1983,Niu1984}. Thouless considered an equilibrium state uniformly occupying all the bands below a Fermi surface, then the charge transported over an adiabatic cycle is equal to the summation of the first Chern numbers of all the occupied bands. This discovery shares the same topological origin as the integer quantum Hall effect\cite{Thouless1982,Xiao2010}. Experimentally, the Thouless pumping was observed in ultracold atoms platforms\cite{Wang2013,Nakajima2016a,Lohse2016}. A generalized Thouless pumping\cite{Wang2015,Zhou2015,Raghava2017} takes into account the interband coherence of the initial state, due to which, besides the topological contribution, an additional component in the adiabatic charge pumping appears. This additional component can be continuously and extensively controlled in experiments\cite{Ma2018}.

In this study, we show that through the generalized Thouless pumping, it is possible to reveal the detailed dynamics of the single-particle interband coherence in the presence of many-body interaction. Starting with the ground state of a non-interacting Hamiltonian $H_i$, we quench one of its parameters to a different value to prepare a non-equilibrium state of the post-quenched Hamiltonian $H_f$. Meanwhile, we also propose to switch on an interacting term in the post-quenched Hamiltonian. Let the state evolve under $H_f$ over a time interval $(0,\tau]$, then we take the time-evolved many-body state as the initial state of an adiabatic charge pumping protocol, which is implemented by  a time-dependent Hamiltonian $H_m(t)$. The many-body time evolution duration $\tau$ will be scanned, thus we can investigate the detailed dynamics of the interband coherence. Here $H_m(t)$ acts as a probing tool used in adiabatic pumping and hence there is no need for us to introduce many-body interaction to $H_m(t)$.   We computationally investigate the integral of the local current flown from one unit cell to its nearest neighbor in periodic boundary condition (PBC) as the amount of pumped charge per cycle.  An important physical insight is the following: for the same initial state that our adiabatic pumping starts with,  the contribution from the interband coherence to the charge pumping dramatically changes by choosing driving protocols with different switch-on rates. Though this piece of physics that is only known recently \cite{Wang2015,Zhou2015,Raghava2017}, it becomes possible to "see"  how single-particle interband coherence is affected by many-body interaction. Interestingly, it is found that the interband coherence can display an oscillating behaviour instead of simply decaying to its thermal value, for a range of $\tau<\tau_c$, where $\tau_c$ is the approximate relaxation time inferred from the evolution of the entanglement entropy and the nearest-neighbor correlation function. As such, the generalized Thouless pumping can be employed to detect the amount of quantum coherence of interacting many-body states during a thermalization process. The proposed scenario also makes it possible to connect experimental observable (pumped charges in different protocols)  with abstract coherence measures. Indeed,  comparing the results of the adiabatic pumping with a coherence measure defined from the single particle density matrix in the sublattice representation (which is not in the band representation), we find that the difference in the amount of pumped charges between different protocols vanishes when the single-particle coherence measure we examined reaches its local minima in regimes of relatively weak interaction.

This paper is organized as follows. In Sec. \ref{model}, a spinless fermionic tight-binding model is introduced, and the pre-quenched, post-quenched and adiabatic pumping Hamiltonian are specified in real space basis. We also briefly derive the expectation value of the current in the adiabatic charge pumping. In Sec. \ref{results}, we present the numerical results of the adiabatic pumping of the system after the quench with and without an interaction term separately, revealing the dependence of the pumped charge on the driving protocols. This is compared with computational studies of a single-particle coherence measure.   Finally, we summarize our results and discuss some motivating questions in Sec.~\ref{conclusion}.

\section{Stage of our analysis: model, adiabatic pumping, entanglement entropy, and coherence measure}\label{model}

\subsection{Model}

The system considered here is the Rice-Mele model\cite{Rice1982} with the following tight-binding lattice Hamiltonian:
\begin{eqnarray}
H_i =&& -\sum_{l=0}^{L-1}[J_i+(-1)^l\delta]\left(c^\dagger_{l}c_{l+1}+
\textrm{H.c.}\right)\nonumber\\
&& +\Delta\sum_{l=0}^{L-1}(-1)^l c^\dagger_{l}c_{l}.
\end{eqnarray}
Here $c_{l}/c^{\dagger}_{l}$ is the spinless fermionic annihilation/creation operator at site $l$. $J_i$ is the hopping amplitude, $\delta$ represents the amount of staggering in the hopping amplitude, and $\Delta$ describes an energy bias between the sublattices. Throughout this paper, we consider the half-filling scenario with periodic boundary condition (PBC), i.e., $c_{L}=c_{0}$. Note that in this system, each unit cell has two sublattices.

\begin{figure}
\centering
\includegraphics[width=1.0\linewidth]{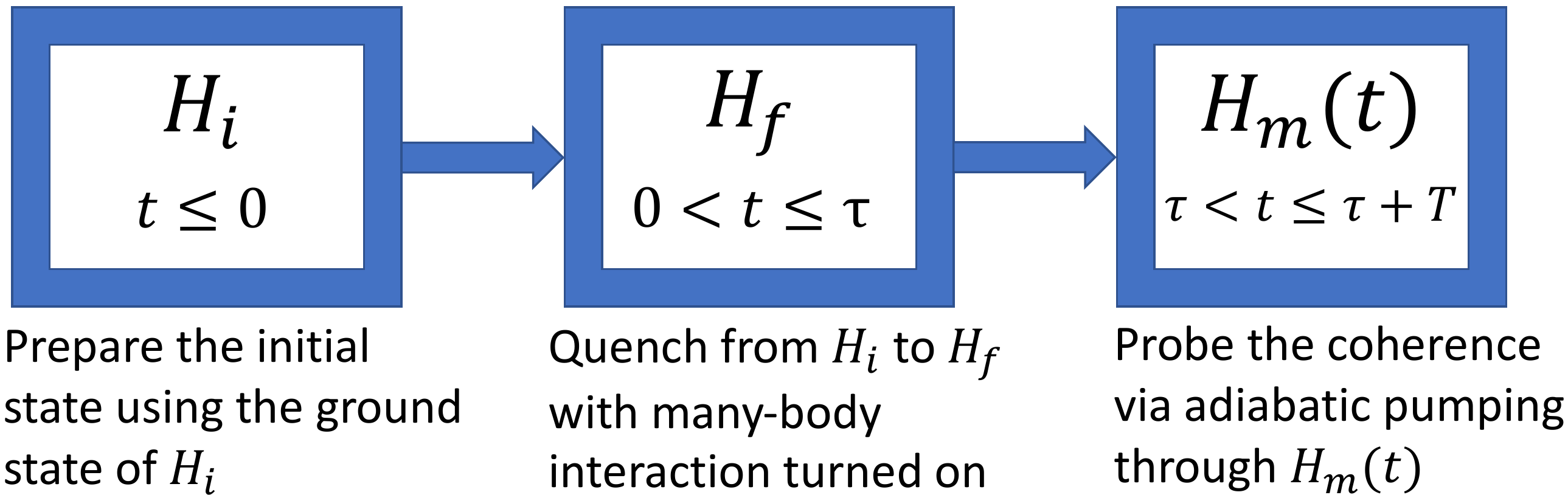}
\caption{Schematic representation of the entire time evolution of the system.}
\label{fig}
\end{figure}

In experiments, a quantum quench can be implemented through a sudden change of one of the Hamiltonian's parameters. Besides, a nearest neighbor interaction in the Hamiltonian, i.e., $U\sum_{l=0}^{L-1}n_{l}n_{l+1}$ where $n_{l}=c^{\dagger}_{l}c_{l}$ is the particle number operator at site $l$, may be also switched on along with the quench. Based on these considerations, we separate the time evolution of the system into three stages, each of which is governed by a different Hamiltonian as schematically shown in Fig.~\ref{fig}. $H_f$ and $H_m(t)$ will be specified shortly.

\begin{equation}
H(t)=\begin{cases}
H_{i} & t\leq 0\\
H_{f} & 0<t\leq\tau\\
H_{m}(t) & \tau<t\leq\tau+T
\end{cases},
\end{equation}

At time $t\leq 0$, the system is in the ground state of $H_i$. At $t=0$, we quench the system's parameters and let it evolve under the post-quenched Hamiltonian $H_f$, given by

\begin{eqnarray}
H_f =&& -\sum_{l=0}^{L-1}[J_f+(-1)^l\delta]\left(c^\dagger_{l}c_{l+1}+
\textrm{H.c.}\right)\nonumber\\
&& +\Delta\sum_{l=0}^{L-1}(-1)^l c^\dagger_{l}c_{l}+U\sum_{l=0}^{L-1}n_{l}n_{l+1},
\end{eqnarray}

where $U$ is the strength of the interaction. After evolving the system under $H_f$ over the time interval $(0, \tau]$, we will obtain a non-equilibrium many-body state possibly with interband coherence. We then take this state to execute an adiabatic charge pumping under another Hamiltonian $H_m(t)$, given by

\begin{eqnarray}
H_m(t) =&& -\sum_{l=0}^{L-1}[J_f+(-1)^l\delta'(t)]\left(c^\dagger_{l}c_{l+1}+
\textrm{H.c.}\right)\nonumber\\
&& +\Delta'(t)\sum_{l=0}^{L-1}(-1)^l c^\dagger_{l}c_{l}.
\end{eqnarray}

Here time dependence is introduced in the parameters $\delta'(t)=R_{\delta}\cos(2\pi \beta(s))$ and $\Delta'(t)=R_{\Delta}\sin(2\pi\beta(s))$, where $R_{\delta,\Delta}>0$. $\beta(s) \in [0,1]$ denotes a function of $s$, which specifies the driving protocol to be studied. $s=\frac{t-\tau}{T}\in [0,1]$ is the scaled time and $T$ is the total time duration of the adiabatic pumping.
Notice that only $H_f$ involves an interacting term, whereas $H_m(t)$ is designed to be non-interacting, in order to preserve the quantum coherence in the adiabatic pumping.

\subsection{Adiabatic pumping}

Over one period $T$, the pumped charge from site $l$ to $l+1$ can be expressed as the integral of the instantaneous local current ${\cal{J}}_l(t)$ at the cross section between site $l$ and site $l+1$,
\begin{equation}
Q_l(T+\tau,\tau) = \int^{T+\tau}_\tau{\cal{J}}_l(t)dt,
\end{equation}
where ${\cal{J}}_l(t) = \langle \psi(t)|\hat{{\cal{J}}}_l(t)|\psi(t)\rangle$ is the expectation value of the local current operator $\hat{{\cal{J}}}_l(t)$. Note that the local current operator is time-dependent because of the parametric temporal dependence in $\delta'(t)$. Explicitly, using the continuity equation,
\begin{equation}
\hat{{\cal{J}}}_l(t)-\hat{{\cal{J}}}_{l+1}(t) = -i[\hat{n}_l,\hat{H}_m(t)],
\end{equation}
we have
\begin{equation}
{\hat{{\cal{J}}}_l(t) = i[J_f+(-1)^l\delta'(t)](c^\dagger_l c_{l-1}-\rm{H.c.}).}
\end{equation}

To obtain the number of pumped charge per cycle, we integrate the expectation value of current over the period $T$,
\begin{equation}
Q_l(T+\tau,\tau) = \int^{T+\tau}_\tau\langle \psi(t)|\hat{{\cal{J}}}_l(t)|\psi(t)\rangle dt,
\end{equation}
with $|\psi(t+\tau)\rangle = U_m(t+\tau,\tau)|\psi(\tau)\rangle$. Here setting $\hbar=1$, $U_m(t+\tau,\tau)={\cal \hat{T}}e^{-i\int_\tau^{t+\tau}H_m(t')dt'}$, and $|\psi(\tau)\rangle = U_f(\tau,0)|\psi(0)\rangle$ with $U_f(\tau,0)=e^{-iH_f\tau}$. Note that $Q_l$ is dependent on the site index, but independent on the unit cell index under PBC, and here we are interested in the intercell current, i.e. current flown from unit cell to its nearest neighbor unit cell. Setting $T$ in the adiabatic limit and using the transnational invariance of the system, we can convert the above equation into the form,
\begin{equation}
Q_l(T+\tau,\tau) = \int_\tau^{T+\tau}dt \sum_k j_l(k,t).\label{Q_l}
\end{equation}
Here, $k$ is the quasi-momentum, and $j_l(k,t)$ is the expectation value of the single particle current operator, defined as
\begin{eqnarray}
j_l(k,t)&&=\sum_{k\in BZ} \langle \phi(k,t)|\hat{j}_l|\phi(k,t)\rangle\nonumber\\
      &&=\frac{1}{N}\sum_{k\in BZ}\langle u(k,t)|\partial_k \hat{{\cal{H}}}(k,t)|u(k,t)\rangle,
\end{eqnarray}
with ${\cal{H}}(k,t)$ being the single particle Hamiltonian in momentum space, $|\phi(k,t)\rangle=|k\rangle\otimes |u(k,t)\rangle$ the single particle Bloch state, and $N$ is the number of particles in the system.

We adopt the reparameterization $s=\frac{t-\tau}{T}$ for ease of notation in this section. A generic initial state in a non-interacting $n$-band model is of the form
\begin{eqnarray}\label{sec2:initial-state}
|u(k,s)\rangle=
\sum_na_{n}(k,s)e^{-iT\int_0^s \varepsilon_{n}(k,s^\prime)\ud s^\prime}|u_{n}(k,s)\rangle,
\end{eqnarray}
where $|u_{n}(k,s)\rangle$ and $\varepsilon_{n}(k,s)$ satisfy the instantaneous eigenvalue equation
\begin{eqnarray}
{\cal H}(k,s)|u_n(k,s)\rangle=\varepsilon_n(k,s)|u_n(k,s)\rangle.
\end{eqnarray}
Here, $\ket{\mu_n(k,s)}$ is assumed to be in a parallel-transport gauge, i.e.,
\begin{eqnarray}
\inp{\mu_n(k,s)}{\dot{\mu}_n(k,s)}=0,
\end{eqnarray}
where the dot denotes the derivative w.r.t.~$s$.

Substituting Eq.~(\ref{sec2:initial-state}) into the time-dependent Schr\"{o}dinger equation
\begin{eqnarray}
i\frac{\partial}{\partial s}\ket{u(k,s)}=T{\cal H}(k,s)\ket{u(k,s)},
\end{eqnarray}
we have
\begin{eqnarray}
&&\dot{a}_m(k,s)=\nonumber\\
&&-\sum_{n\neq m}
a_n(k,s)e^{iT\int_0^s\omega_{mn}(k,s^\prime)\ud s^\prime}
\inp{u_m(k,s)}{\dot{u}_n(k,s)},\nonumber\\
\end{eqnarray}
where $\omega_{mn}(k,s)\equiv\varepsilon_m(k,s)-\varepsilon_n(k,s)$.
Solving this equations with adiabatic perturbation theory yields
\begin{eqnarray}
a_m(k,s)=a_m(k,0)+\frac{1}{T}\sum_{n\neq m}a_n(k,0)W_{mn}(k,s^\prime)|_{0}^{s},\nonumber\\
\label{a_m}
\end{eqnarray}

with
\begin{eqnarray}
W_{mn}(k,s)=i\frac{\langle u_m(k,s)|\dot{u}_n(k,s)\rangle}{\omega_{mn}(k,s)}
e^{iT\int_0^s\omega_{mn}(k,s^\prime)\ud s^\prime}.\nonumber\\
\end{eqnarray}
For later reference, note that
\begin{eqnarray}
W_{mn}(k,s) &\propto & \langle u_m(k,s)|\dot{u}_n(k,s)\rangle \nonumber \\
& =  & \langle u_m(k,s(\beta))| \frac{d u_n(k,s(\beta))}{d\beta}\rangle \frac{d \beta}{d s}.  \label{rateibc}
\end{eqnarray}

Now, inserting Eq.~(\ref{a_m}) into  Eq.~(\ref{Q_l}), we obtain, after some algebra \cite{Wang2015,Zhou2015},
\begin{eqnarray}
Q_l(T+\tau,\tau)=Q_{\textrm{TP}}+Q_{\textrm{IM}}+Q_{\textrm{NG}}+Q_{\textrm{IBC}},
\end{eqnarray}
with
\begin{eqnarray}
Q_{\textrm{TP}} = \frac{1}{2\pi}\sum_n\int_{-\pi}^{\pi}dk\int_0^{1}ds \rho_{n}(k,0)\Omega_{n}(k,s),
\end{eqnarray}
\begin{eqnarray}
Q_{\textrm{IM}}=\frac{T}{2\pi}\sum_n\int_{-\pi}^\pi dk\int_{0}^{1}ds \rho_n(k,0)\frac{\partial \varepsilon_n(k,s)}{\partial k},
\end{eqnarray}
\begin{eqnarray}
Q_{\textrm{NG}}=&&\frac{1}{\pi}\int_{-\pi}^\pi dk\int_0^1 ds\sum_{m<n}\textrm{Im}[a_m^*(k,0)a_n(k,0)\nonumber\\
&&\times\inp{u_m(k,s)}{\frac{\partial}{\partial k}u_n(k,s)}],
\end{eqnarray}
\begin{eqnarray}
Q_{\textrm{IBC}}=&& -\frac{1}{\pi}\sum_{m\neq n}\int_{-\pi}^{\pi}dk\int_0^1 ds \frac{\partial \varepsilon_{m}(k,s)}{\partial k}\nonumber\\
&&\times\mathrm{Re}[a_{m}^*(k,0)a_{n}(k,0)W_{mn}(k,0)].
\label{q_ibc}
\end{eqnarray}
Here, $m,n$ are energy band indices, $\rho_{n}(k,0)$ is the population of the $n$-th energy band, and $\Omega_{n}(k,s)$ is its Berry curvature, defined as
\begin{equation}
\Omega_{n}(k,s)=i\langle\partial_{s}u_{n}(k,s)|\partial_{k}u_{n}(k,s)\rangle+{\rm c.c.}
\end{equation}

Clearly, $Q_{\textrm{TP}}$ is a weighted sum of integrals of Berry curvature and therefore has a topological origin. Furthermore, $Q_{\textrm{IM}}$ and $Q_{\rm NG}$ are independent of the switch-on rate of the adiabatic driving field, whereas $Q_{\textrm{IBC}}$ is sensitive to it, since $W_{mn}(k,0)$ is related to the switch-on rate $\frac{d\beta}{ds}|_{s=0}$ at the start of the adiabatic protocol [see Eq.~(\ref{rateibc})].  Note also that the term $Q_{\textrm{IBC}}$ manifests certain quantum coherence in the energy band representation averaged over the quasimomentum $k$, as it contains the off-diagonal density matrix elements  in this representation at each individual values of $k$.

\begin{table}
\caption{Driving protocols of the adiabatic pumping and their switching-on rates}
\centering
\label{tab:1}
\begin{tabular}{ |c|c| }
\hline
Protocols & Switch-on Rates \\
\hline
$\beta_C(s) = 1-\cos(\pi s/2)$ & 0 \\
\hline
$\beta_L(s) = s$ & 1 \\
\hline
$\beta_S(s) = \sin(\pi s/2)$ & $\pi/2$ \\
\hline
$\beta_F(s) =(\sqrt{s+\frac{1}{25}}-\frac{1}{5})/(\sqrt{1+\frac{1}{25}}-\frac{1}{5})$ & $(\sqrt{1+25}+1)/2$ \\
\hline
\end{tabular}
\end{table}

\subsection{Entanglement entropy and coherence measure}

To monitor the dynamics of coherence in the presence of many-body interaction, we may examine the entanglement entropy between the left half and the right half of the system. It is defined as
\begin{eqnarray}
S_{\rm ent}(t) = -\frac{1}{|A|}\mathrm{tr}_A[{\rho}_A(t)\mathrm{log}\rho_A(t)].
\end{eqnarray}
The subsystem $A$ is the left half of the system, with $|A|=L/2$ and the reduced density matrix $\rho_A(t)=\mathrm{tr}_{A^{\rm c}}|\psi(t)\rangle\langle\psi(t)|$. On the other hand, note that the resource theory of coherence has been put forward, based on which a number of coherence measures has been proposed \cite{Baumgratz2014}. To quantify the single-particle interband coherence, we resort to the $l_1$ norm of coherence, defined as
\begin{eqnarray}
C(t) = \frac{1}{N}\sum_k\sum_{\mu \neq \nu}|\rho_{\mu\nu}(k,t)|+|\rho_{\nu\mu}(k,t)|.
\label{CM}
\end{eqnarray}
Here, $\rho_{\mu\nu}(k,t)$ is
the single-particle density matrix (SPDM), expressed as
\begin{equation}
\rho_{\mu\nu}(k,t) = \langle\psi(t)|c^{\dagger}_{\mu,k} c_{\nu,k}|\psi(t)\rangle,
\end{equation}
where $c^\dagger_{\mu,k}$ is the fermionic creation operator at quasi-momentum $k$ for the sublattice site indexed by $\mu$ in each unit cell. It is worth noting that the $l_1$ norm of coherence can be estimated in experiments by other means \cite{Zhang2018}. Here, we shall connect this coherence measure with charge pumping which is also experimentally accessible in a rather direct way.

\section{Results and Discussions}\label{results}

\begin{figure}
\centering
\includegraphics[width=1.0\linewidth]{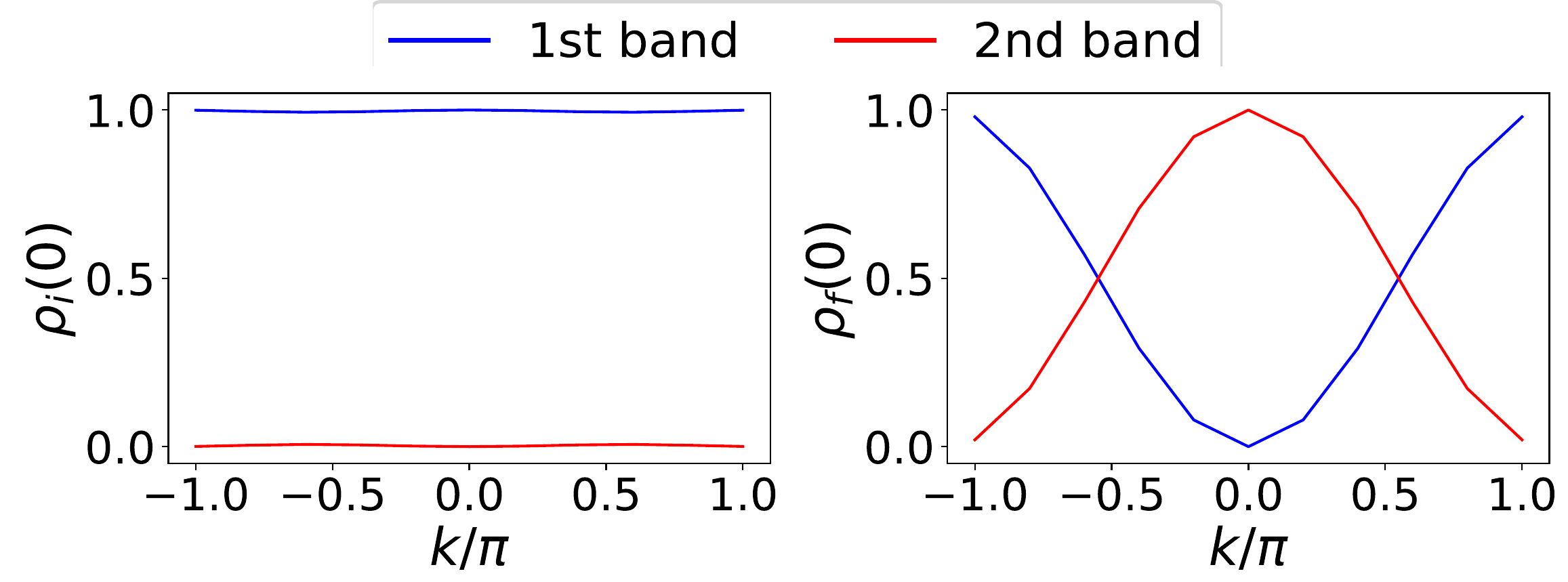}
\caption{Left/Right panel: probability distribution $\rho_i(0)/\rho_f(0)$ on two bands along the quasi-momentum of the initial state before/after quenching of systems. Parameters adopted are $\delta = 0.85$, $\Delta = 0$, and $J_i=1$/$J_f=-1$.}
\label{fig0}
\end{figure}

In this section, we numerically simulate the adiabatic charge pumping in our system, and the time evolution of the many-body states is computed with the Python package QuSpin\cite{QuSpin}. We initialize the state of our system at the many-body ground state $|\psi_i\rangle$ of the pre-quenched Hamiltonian $H_i$, with system parameters $J_i=1, \delta = 0.85$, $\Delta = 0$, and $L=14$ at half-filling. To introduce interband coherence into the initial state, we quench one of the system's parameters $J$ suddenly from $J_i=1$ to a different value, say, $J_f=-1$ in $H_f$. In the basis of the post-quenched Hamiltonian $H_f$, the state $|\psi_i\rangle$ populates both the valence and conduction bands. In Fig.~\ref{fig0}, we show the probability distribution $\rho(t=0)$ of the initial state on the two bands of the pre-quenched and post-quenched Hamiltonians as a function of the quasi-momentum $k$.

Starting from $\ket{\psi_i}$, the evolving state undergoes the evolution governed by $H_f$ for a time interval $(0,\tau]$. Finally, the evolved state $|\psi(\tau)\rangle=U_f(\tau,0)|\psi_i\rangle$ is adiabatically pumped under $H_m(t)$ with $R_{\delta}=0.85$ and $R_{\Delta}=2$. To study the influence of switch-on rates on the adiabatic pumping, we adopt four different driving protocols $\beta(s)$, as summarized in Table \ref{tab:1}. The protocol $\beta_F(s)$ has the largest switch-on rate. Therefore it should introduce the largest $Q_{\textrm{IBC}}$ to the pumping charge. On the contrary, the protocol $\beta_C(s)$ has a vanishing switch-on rate. Thus $Q_{\textrm{IBC}}$ should be vanished for this protocol. In the following, both non-interaction quench ($U=0$) and interaction quench ($U\neq 0$) will be investigated. We will show that, apart from the non-interaction terms in $H_f$, the effect of many-body interaction on the dynamics of the interband coherence can be revealed through adiabatic pumping by varying $\tau$.

\subsection{Non-interaction Quench, $U=0$}

\begin{figure}
\centering
\includegraphics[width=1.0\linewidth]{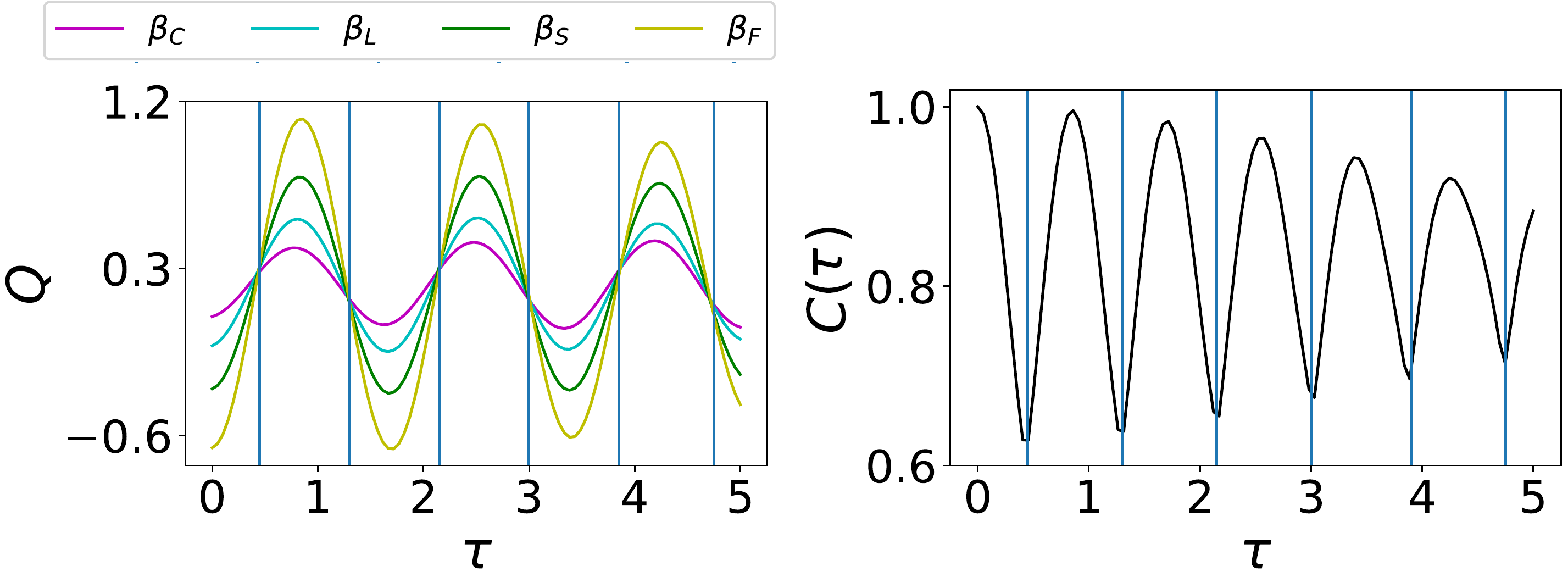}
\caption{Left panel: numerical results of the adiabatic pumping with different driving protocols. Right panel: dynamical behavior of the coherence measure in the sublattice representation. Here parameters are set as $J_f=-1$, $R_{\delta}=0.85$ , $R_{\Delta}=2$ in $H_m(t)$ and $J_f=-1$, $\delta=0.85$, $\Delta=0$, $U=0$ in $H_f$. }
\label{fig1}
\end{figure}

Setting $U=0$ in $H_f$, here we present the numerical results of the pumped charge $Q$ vs. $\tau$,  with different driving protocols plotted in the left panel of Fig.~\ref{fig1}. We observe that at different $\tau$, the interband coherence of the many-body state is manifestly revealed by the differences in the amount of pumped charges between different driving protocols with a varying switch-on rate. Roughly speaking, the difference in the charge pumped caused by the switch-on rate of the adiabatic protocol acts as a strong witness of the interband coherence in the system. That is, one driving protocol alone cannot directly reveal the coherence effect explicitly, but two or more driving protocols with different switch-on rates will directly tell if there is considerable interband coherence effect left in the system. Remarkably, from our computational studies we observe that there are some special points (in terms of $\tau$)  at which different driving protocols yield the same amount of pumped charges. That is,  if we let the system evolve for such $\tau$ values, the contribution from the interband coherence is vanishing.  It can be inferred that in the absence of any many-body interaction, the contribution from interband coherence effect to adiabatic pumping is oscillating. This is consistent with the expression of $Q_{\rm IBC}$ shown above, which contains the off-diagonal matrix elements that are expected to be oscillating with the free evolution time $\tau$.

To connect the physics of interband coherence as manifested by these special time points in the duration of evolution, we study the dynamical behavior of a different single-particle coherence measure of the state, as defined in Eq.~(\ref{CM}).  We find that these special time values correspond to the local minima of this second coherence measure that depicts the coherence between two sublattice sites, as can be seen in the right panel of Fig.~\ref{fig1}. Note that the oscillation behaviour of the results again resembles to the Rabi oscillation in a non-interacting two-level system. In fact, when $U=0$, the quasi-momentum $k$ is a good quantum number under PBC. We can interpret the system as a collection of independent two-level systems at each $k$ point. With a proper choice of parameters, we obtain two bands with small curvature and nearly uniform band gap along the quasi-momentum $k$. Thus it can be expected quantum coherence for different $k$ in the sublattice representation oscillates at almost the same frequency and the oscillation period is indeed consistent with the band gap of our system as well as the oscillation period of $Q_{\rm IBC}$.   This understanding also indicates that, in the presence of bands with large curvature, the interband coherence contribution to pumping, $Q_{\rm IBC}$, and the coherence measure $C(t)$ (both as certain $k$-averaged quantities) will display certain damped oscillations even without any interaction effect or decoherence effect (that is, the phases of oscillations in the off-diagonal elements in energy-band representation and in the sublattice representation will be scrambled). To distinguish this effect as much as possible from the many-body interaction effect we aim to examine, we have attempted to optimize the system's parameters in order to reach a regime where the two bands of $H_f$ are nearly flat.  It is due to this subtle treatment, that the oscillation amplitudes of $C(\tau)$ and the interband coherence contribution to adiabatic pumping do not appreciably decay, as shown in Fig.~\ref{fig1}.    This feature will be compared with cases if we switch on the many-body interaction in $H_f$.

\subsection{Interaction Quench, $U\neq 0$}

Besides quenching one of the parameters in the Hamiltonian, we can also study quenches in the presence of interactions $U \neq 0$. To do this, we consider the quench with $U \leq 2$ and expect the system to thermalize under the many-body evolution Hamiltonian $H_f$ during the time interval $(0,\tau]$, i.e., the non-equilibrium initial state evolves towards an equilibrium distribution. In this thermalization process, memories of the initial state tend to be washed out gradually, and the final state is expected to be a featureless thermal state with little interband coherence left.  We can monitor this thermalization process with the entanglement entropy and the nearest-neighbor correlation function.

\begin{figure}
\centering
\includegraphics[width=1.0\linewidth]{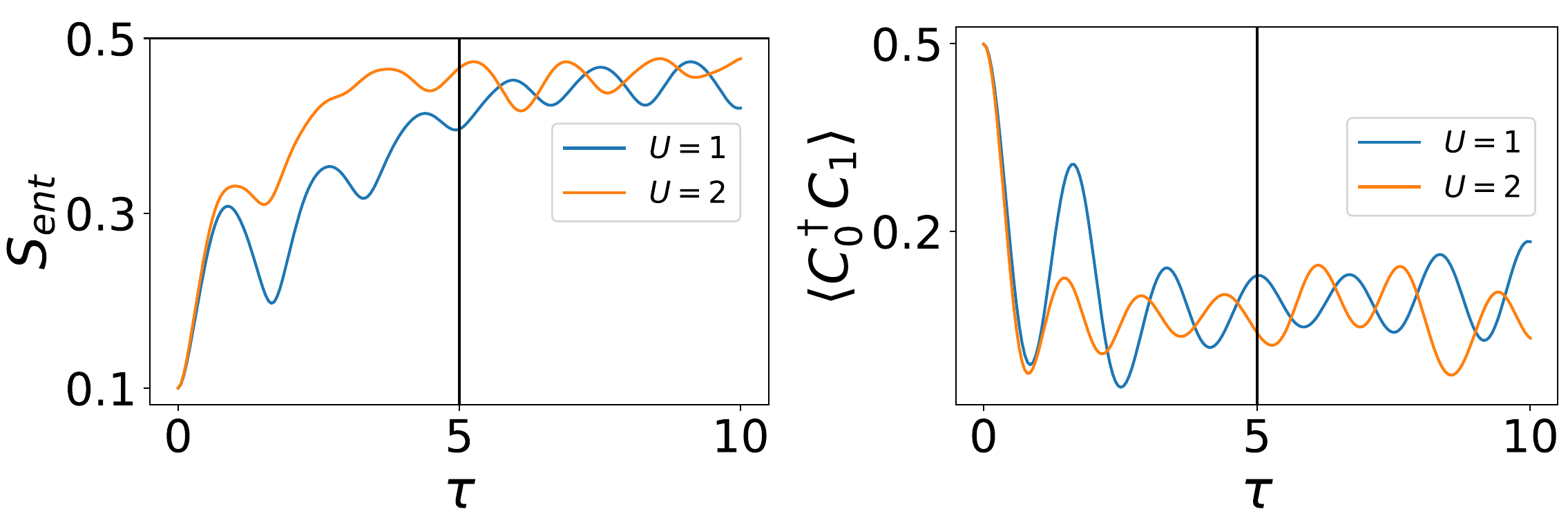}
\caption{Left panel: entanglement entropy density between the left half part and the right half part of the system as a function of time $\tau$ for different interaction strength $U$. Right panel: short time behavior of the nearest-neighbor correlation functions for different interaction strength $U$ in $H_f$. Other parameters adopted are $J_f=-1$, $\delta=0.85$, $\Delta=0$ in $H_f$. }
\label{fig2}
\end{figure}

\begin{figure}
\centering
\includegraphics[width=1.0\linewidth]{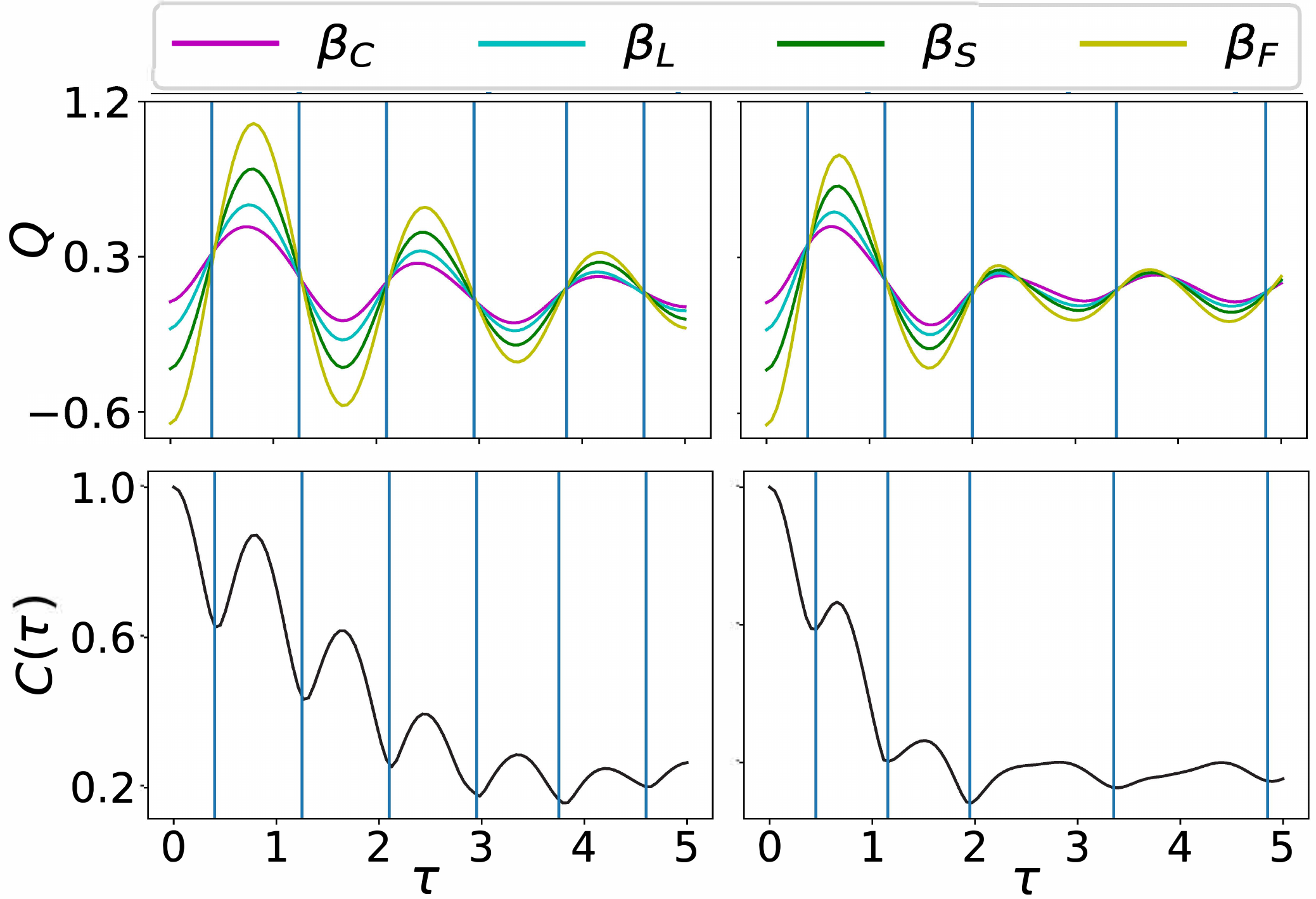}
\caption{Top Left/Right panel: numerical results of the adiabatic pumping with different driving protocols with $U=1/U=2$ in $H_f$. Bottom Left/Right panel: dynamical behavior of the coherence measure in the sublattice representation with $U=1/U=2$ in $H_f$. Other parameters set here are $J_f=-1$, $R_{\delta}=0.85$ , $R_{\Delta}=2$ in $H_m(t)$ and $J_f=-1$, $\delta=0.85$, $\Delta=0$ in $H_f$.}
\label{fig3}
\end{figure}

In Fig.~\ref{fig2}, we present  the entanglement entropy density between the left half and the right half parts of the system, and the correlation function as a function of time $\tau$ for different interaction strength $U$. From the behaviors of these two functions, we deduce that it should be sufficient to consider the time interval $(0,\tau]$ with $\tau\leq5$, as the entanglement entropy is almost saturated and the nearest-neighbor correlation function is roughly steady for $\tau\geq5$ (with some revivals however).  The numerical results of the adiabatic charge pumping with different pumping protocols and the dynamical behavior of the coherence measure of the many-body states with $U=1$ and $U=2$ in $H_f$ are presented in Fig.~\ref{fig3}.

From Fig.~\ref{fig3}, it follows that the overall differences in the pumped charges between different pumping  protocols are appreciably decreasing with the duration of many-body interaction time $\tau$. This is in clear contrast to the non-interaction quench scenario where the differences between different pumping protocols remain significant.  Thus, the charge pumping differences induced by different switch-on rates of the various pumping protocols do a good job in manifesting the dynamics of quantum interband coherence.  With this understanding, results in the upper panels of  Fig.~\ref{fig3} indicates a clear decreasing trend in the differences between different pumping protocols, thus showing clearly an overall decrease in the underlying single-particle interband coherence in the system as $\tau$ increases. These results agree well with the time dependence of the different coherence measure $C(t)$ presented in the bottom panels in Fig.~\ref{fig3}, where $C(t)$ is also seen to decrease with many-body interaction time. Physically, each individual particle in the system is in the presence of a bath composed of other interacting particles and as a consequence, the off-diagonal elements of the SPDM are expected to decay as a characteristic feature of thermalization. For sufficiently long time of many-body interaction, contributions from single-particle interband coherence to charge pumped are expected to diminish, as also seen from the upper panels of Fig.~\ref{fig3}, where almost all the considered pumping protocols produce the same result.
Furthermore, both the coherence measure (bottom panels of Fig.~\ref{fig3}) and the pumping charge differences between different protocols (upper panels of Fig.~\ref{fig3}) continue to exhibit an oscillatory behaviour on top of their overall decay. That is, the oscillatory behavior of the single-particle interband coherence effect gives rise to similar observable effects in the adiabatic pumping outcome with different protocols. This strengthens the view that adiabatic pumping with initial state coherence can be exploited to study detailed dynamics of the quantum coherence.
In addition, we can also examine the first few local minima of the coherence measure vs. the time points at which different pumping protocols give the same results. It is observed that for $U=1$, there is an excellent correspondence of the local minima of the coherence measure and the time points at which different driving protocols give the same result, as shown in the left panels of Fig.~\ref{fig3}. However, in the right panel of Fig.~\ref{fig3} for $U=2$, this interesting correspondence can still be seen but it is not as beautiful as in the case of $U=1$.  Because adiabatic pumping is affected by interband coherence (hence coherence in the energy-band representation) and the coherence measure $C(t)$ depicts certain coherence in sublattice representation, we do not expect after all a quantitatively simple relation between $C(t)$ and the interband coherence effect in adiabatic pumping.

\section{Conclusion}\label{conclusion}

In conclusion, we have found that the detailed dynamics of single-particle interband coherence in the presence of many-body interaction can be revealed through adiabatic charge pumping with different pumping protocols. This is possible because the contributions of the interband coherence to adiabatic pumping can be adjusted by considering different pumping protocols with different switch-on rates. Thus, quantum coherence can not only be characterized by some theoretical measures, it can be also witnessed by physical observables (pumped charges in our case here).
Of particular interest and enhancing this claim, we find that even the oscillating behavior of one quantum coherence measure is well echoed by the oscillating differences in adiabatic pumping between different protocols.
It is also observed that, at least in the regime of relatively weak interaction strength, when the coherence measure reaches its local minima as we scan the many-body interaction time, all different pumping protocols give the same adiabatic pumping results. This offers an experimentally feasible means to track the abstract coherence measure with the aid of adiabatic charge pumping. We also note that the dynamics of the $k$-averaged coherence measure based on SPDM in sublattice representation can only be witnessed well if the bands are relatively flat.
Finally, we have not explored regimes of strong interaction because other physics might emerge there. It may be also interesting if one considers disorder quenches where thermalization is not guaranteed and many-body localization (MBL) may emerge. If there is MBL, then the dynamics of interband coherence with MBL would be also a promising topic.

\begin{acknowledgments}

J.G.~is supported by the Singapore NRF Grant No.~NRF-NRFI2017-04 (WBS No.~R-144-000-378-281). D.-J.Z.~acknowledges support from the National Natural Science Foundation of China through Grant No.~11705105 before he joined NUS. L.Z. acknowledges support from the Young Talents Project at Ocean University of China (Grant No.~861801013196 and 841912009).

\end{acknowledgments}


%

\end{document}